\documentclass[a4paper,12pt]{article}
\usepackage[latin1]{inputenc}	%% permet d'utiliser les accents
\usepackage{graphicx}
\usepackage{amsfonts}
\usepackage{amssymb}

\usepackage{graphics}
\newcommand{\Rbb}{\mathbb{R}}

\newcommand{\Leb}[2]{
  \ifnum #2=1
  L^#1(\Rbb,\,\ud x)
  \fi
  \ifnum #2=2
  L^#1(\Rbb^2,\,\ud^2 \vec{x}\,)
  \fi}

\newcommand{\ud}{\mathrm{d}}

\title{Some examples of exponentially harmonic maps}
\author{A  D Kanfon$^{1,2}$,
A  F\"uzfa$^{2,3}$ and
D Lambert$^{3}$ }

\begin{document}

\maketitle
\begin{small}   ${}^{1}$Universit\'e d'Abomey-Calavi, Unit\'e de recherche en physique th\'eorique, BP 2628, Porto-Novo, B\'enin

 ${}^{2}$Universit\'e Catholique de Louvain-la-Neuve, Unit\'e FYMA, chemin du cyclotron, 2, B-1348 Louvain-la-Neuve, Belgium

 ${}^{3}$Facult\'es Universitaires N.-D. de la paix, Unit\'e de systemes dynamiques, Rue de Bruxelles, 61, B-5000 Namur, Belgium\\

E-mail: kanfon@yahoo.fr, afu@math.fundp.ac.be and d.lambert@fundp.ac.be    \end{small}

\begin{abstract}
 The aim of this paper is to study some examples of exponentially harmonic maps. We study such maps
firstly on flat euclidean and Minkowski spaces and secondly on Friedmann-Lema\^ itre universes.
We also consider some new models of exponentially harmonic maps which are coupled with gravity
which happen to be based on a generalization of the lagrangian for bosonic strings coupled with
dilatonic field.
%\vfill
%\begin{flushright}
%UCL-IPT--02--08\\
%May 2002
%\end{flushright}

\end{abstract}
\section{Introduction}

 Exponentially harmonic maps were introduced by James Eells and studied by  J. Eells
and L. Lemaire \cite{EeL1}. These maps generalize usual harmonic maps \cite{EeL2} in the following
sense. Let $(M,g)$ and $(N, h)$ be two Riemannian manifolds and $\phi: M\rightarrow N: x\rightarrow\phi(x)$,
a smooth map. An exponentially harmonic map is then an extremal of the following functional:
\begin{equation}\label{functional}
  E(\phi)=\int_{M} \exp(e(\phi)) d\mu(\phi)
\end{equation}
 where $d\mu(\phi)$ is the riemannian volume element  and
\begin{equation}\label{Riemannian volume el}
 e(\phi)=\frac{1}{2}\frac{\partial \phi^a}{\partial x^{\mu}} \frac{\partial \phi^b}{\partial x^{\nu}} g^{\mu\nu}h_{ab}
\end{equation}
is the so called ``energy density'' of the map $\phi$. In local coordinates $(x^{\mu})$ and $(\phi^{a})$,
the equation for exponentially harmonic maps, which derives from the variation of the functional (\ref{functional}),
is

\begin{eqnarray}\label{variation of the functional}
& &\exp(e(\phi))\left\lbrace  g^{\alpha\beta} \left(\frac{\partial^2 \phi^a}{\partial x^\alpha \partial x^\beta}- {\Gamma}^{\gamma(M)}_{\alpha\beta}\frac{\partial \phi^a}{\partial x^{\gamma}}+ \Gamma^{a(N)}_{bc} \frac{\partial \phi^b}{\partial x^\alpha}\frac{\partial \phi^c}{\partial x^\beta}\right)\right.\nonumber\\
& &\left.+ g^{\alpha\mu}g^{\beta\nu}h_{bc}\frac{\partial \phi^a}{\partial x^{\mu}}\frac{\partial \phi^c}{\partial x^{\nu}}  \frac{\partial^2 \phi^b}{\partial x^{\alpha} \partial x^{\beta}}  - g^{\alpha\mu}g^{\beta\nu}h_{bc}  \Gamma^{\gamma (M)}_{\alpha\beta}  \frac{\partial \phi^a}{\partial x^{\mu}}\frac{\partial \phi^b}{\partial x^{\nu}}  \frac{\partial \phi^c}{\partial x^{\gamma}}  \right.\nonumber\\
& &\left.+ g^{\alpha\beta}g^{\mu\nu}h_{bc} \Gamma^{b(M)}_{de}  \frac{\partial \phi^d}{\partial x^{\alpha}}\frac{\partial \phi^e}{\partial x^{\mu}}  \frac{\partial \phi^c}{\partial x^{\nu}}\frac{\partial \phi^a}{\partial x^{\beta}}  \right\rbrace =0
\end{eqnarray}
where $\Gamma^{\gamma (M)}_{\alpha\beta}$  and $\Gamma^{a(N)}_{bc}$ are the Christophel symbols of the Levi-Civita
connection on $M$ and $N$.
This equation involves, as a particular case, the equation of usual harmonic maps. If we drop the $\exp(e(\phi))$
factor and restrict ourselves to the first three terms we get:
\begin{eqnarray}\label{restriction}
g^{\alpha\beta} \left(\frac{\partial^2 \phi^a}{\partial x^\alpha \partial x^\beta}- {\Gamma}^{\gamma(M)}_{\alpha\beta}\frac{\partial \phi^a}{\partial x^{\gamma}}+ \Gamma^{a(N)}_{bc} \frac{\partial \phi^b}{\partial x^\alpha}\frac{\partial \phi^c}{\partial x^\beta}\right)=0
\end{eqnarray}
which is the field equation for a usual harmonic map (which is nothing but a non-linear sigma model in
 the physicist's language). When $N=\mathbb{R}$,
(\ref{restriction}) is simply written:
\begin{eqnarray}\label{N=R}
 \frac{1}{\sqrt{g}}\frac{\partial }{\partial x^{\alpha}}\left( \sqrt{g}g^{\alpha\beta}\frac{\partial \phi}{\partial x^{\beta}}\right) =0
\end{eqnarray}
which is the Laplace equation in local coordinates. Following the work of Luc Lemaire, it is important
to note, however, that the properties of the exponentially harmonic maps are very different of those of usual harmonic maps.
This comes from the fact that the functional  (\ref{functional})  changes completely when we perform a conformal shift on
 the metric $h$: $ h\rightarrow \lambda.h$. Then  (\ref{functional}) becomes:
\begin{equation}\label{conformal shift}
  E_{\lambda}(\phi)=\int_{M} [\exp(e(\phi))]^{\lambda} d\mu(\phi)~~~~~~~~~~~~~\lambda = \mbox{constant}.
\end{equation}
In the case of usual harmonic maps this metric shift has no influence on the harmonic maps equations, the functional is simply multiplied by a constant.

In the sequel we will study somme particular cases of (\ref{variation of the functional}) even when $M$
 is noncompact and (\ref{functional}) unbounded. In this case, equation (\ref{variation of the functional})
 is taken as the definition of what we call an exponentially harmonic map.
 Following Eells and Lemaire, we will consider the energy-momentum tensor asociated to $\phi$:
\begin{eqnarray} \label{energy momentum tensor}
T_{\mu\nu}(\phi)= \exp(e(\phi))\left(g_{\mu\nu}-\frac{\partial \phi^a}{\partial x^\mu} \frac{\partial \phi^b}{\partial x^\nu} h_{ab}\right)
\end{eqnarray}
which is conserved ($g^{\alpha\beta}\nabla_{\alpha} T_{\alpha\beta} (\phi)=0$) when $\phi$ is a
solution of (\ref{variation of the functional}).

It is worth noting that some particular  cases of (\ref{variation of the functional}) were studied
 by mathematicians in the context of the theory of elliptic partial differential equations. Gilbarg
 and Trudinger \cite{GiT} quote, for example, the equation:
 \begin{eqnarray}\label{Trudinger eq}
  \Delta\phi+\beta \sum_{i,j=1}^{n} \frac{\partial \phi}{\partial x^{i}}\frac{\partial \phi}{\partial x^{j}}\frac{\partial^2 \phi}{\partial x^{i}\partial x^j}=0
  ~~~~~~~~~~\phi: U\subset \mathbb{R}^n\rightarrow \mathbb{R}
\end{eqnarray}
deriving from the variational principle: $\delta\int_\cup \exp(\frac{\beta}{2}\sum_{i=1}^{n}(\frac{\partial \phi}{\partial x^{i}})^2)dx^1...dx^n$

Lut us remark finally that, if we want to make some contact with physics, we have to modify  (\ref{functional})  as follows:
\begin{eqnarray}\label{modified functional}
 E'_\lambda&=&\int_M (\exp(\lambda e(\phi)) -1)d\mu(\phi)\nonumber\\
 &\approx&\lambda\int_M \left[ (\frac{1}{2}\partial_\mu \phi^a \partial^\mu \phi_a)+\frac{\lambda}{2}(\frac{1}{2}\partial_\mu \phi^a \partial^\mu \phi_a)^2+...\right]d\mu(\phi)
\end{eqnarray}
which, if we interpret $(\phi^a)$ as a spin vector, looks like  the Hamiltonian for the continuum
limit of a Heisenberg model. When $\lambda$ is small enough, the variational principle
$\delta E'_\lambda=0$ leads to equations which approximate those of the usual harmonic maps
(sigma models). The equations derived from $E'_\lambda$ can be obtained directly from (\ref{variation
 of the functional}) via the shift $\phi \rightarrow \sqrt{\lambda}\phi ~~~~(\lambda > 0)$.

\section{Exponential harmonic maps on flat spaces}
In order to get some feeling about the solutions of (\ref{variation of the functional}), we
study here scalar exponentially harmonic maps on two-dimensional  Euclidean and Minkowskian
manifolds:
$$ \phi:~~E \rightarrow \mathbb{R}~~~~~~~E=\mathbb{R}^2 ~~or~~\mathbb{R}^{1,1}.$$

\subsection{$E=\mathbb{R}^2$}

Let $(x^\mu) =(x,y)$,  $\phi_{x^\mu}= \partial \phi/\partial x^\mu$ and
$\phi_{x^\mu x^\nu}= \partial^2 \phi/\partial x^\mu \partial x^\nu$. Then, equation  (\ref{variation
of the functional}) becomes
\begin{eqnarray}\label{variation of the functional when E = R2}
 (1+\phi_x^2)\phi_{xx}+2\phi_x\phi_y\phi_{xy}+(1+\phi_y^2)\phi_{yy}=0.
\end{eqnarray}
We are searching solutions of the following form: $\phi(x,y)= F(x)+G(y)$. We are thus led to
the equations:
\begin{eqnarray}\label{following_of_the_variation_of_the_functional_when_ER2}
 [1+ (F_x)^2]F_{xx}=-[1+(G_y)^2]G_{yy}=l
\end{eqnarray}
where $l$ happens to be an arbitrary positive real constant. $l=0$ leads, of course,
 to trivial solutions $\phi(x,y)= ax+by$. If $l>0$, let $p=F_x$. This leads to the system:
\begin{eqnarray}
\label{syteme1eq1}
&& F_x=p(x)\\
\label{system1eq2}
&&p_x=\frac{l}{1+p^2}.
\end{eqnarray}
Then (\ref{system1eq2}) gives
\begin{equation}\label{solution1 du syteme1}
 x=\frac{1}{l}(\frac{1}{3} p^3+p-c);~~~c = ~~\mbox{constant}.
\end{equation}
Let us derive (\ref{solution1 du syteme1}) with respect to $F$. This gives (with $k$=constant):
\begin{eqnarray}\label{solution2 du syteme1}
\frac{dx}{dF}=\frac{1}{F_x}=\frac{1}{l}(1+p^2)\frac{dp}{dF} ~~\mbox{and}~~ p^4+2p^2-4l(F+k)=0.
\end{eqnarray}
The elimination of $p$ from   (\ref{solution1 du syteme1}) and  (\ref{solution2 du syteme1})  gives:
\begin{eqnarray} \label{solution3 du syteme1}
F(x)&=&\frac{1}{4l}\left\lbrace [S_+(x;l;c)+S_-(x;l;c)]^4\right.\nonumber\\
& &\left.+2[S_+(x;l;c)+S_-(x;l;c)]^2\right\rbrace -k\\
p &=& S_+(x;l;c)+S_-(x;l;c).
\end{eqnarray}
where $S_{\pm}(x;l;c)=\left\lbrace \frac{3}{2}(c+l x)\pm (1+\frac{9}{4}(c+l x)^2)^{1/2} \right\rbrace^{1/3}$.
Similary we get the solution $G(y)$. A solution of (\ref{variation of the functional when E = R2}) can thus
be written, with the above notations as,
\begin{eqnarray*}
G(y)&=&-\frac{1}{4l}\left\lbrace [S_+(x;-l;c)+S_-(x;-l;c)]^4 \right.\\
& &\left.+2[S_+(x;-l;c)+S_-(x;-l;c)]^2\right\rbrace -k
\end{eqnarray*}
which leads to the solution we look for.

If we write $q(y)= S_+(x;-l;c)+S_-(x;-l;c)$, then we can represent $\phi(x,y)$ in the
 following parametric form:
\begin{eqnarray*}
& & x=\frac{1}{l}(\frac{p^3}{3}+p-c),~~~y= \frac{-1}{l}(\frac{q^3}{3}+q-c'),\\
 & &\phi(x,y)=\frac{1}{4l}(p^4+2p^2-q^4-2q^2)+cst.
\end{eqnarray*}
It is also possible to solve (\ref{variation of the functional when E = R2}) by the so-called hodograph method
\cite{CoH}. Let $u=\phi_x, v=\phi_y$ and $x=x(u,v),y=y(u,v)$. If $J=x_uy_v-x_vy_u \neq 0$, we get immediatly:
 \begin{eqnarray*}
&& u_x = y_v/J,~~~~~~v_x=-y_u/J\\
&&u_y=-x_v/J,~~~~~v_y=x_u/J
\end{eqnarray*}
We know that if $v_x=u_y$ then $x_v=y_u$. Thus, there exists a function $f(u,v)$ such that $x=f_u$ and $y=f_v$.
Equation  (\ref{variation of the functional when E = R2}) can be written
\begin{eqnarray}\label{1.12}
 (1+u^2)f_{vv}-2uvf_{(uv)}+(1+v^2)f_{uv}=0;
\end{eqnarray}
 or, using polar coordinates, $u=r\cos\theta, v=r\sin\theta$:
 \begin{eqnarray}\label{1.13}
f_{rr}+(r+\frac{1}{r})f_r+(1+\frac{1}{r^2})f_{\theta\theta}=0
\end{eqnarray}
which can be solved  by factorizing: $f(r,\theta)=  R(r)T(\theta)$. We have now to solve the two
equations (with an arbitrary constant $a$):
\begin{eqnarray}
\label{1.14}&& R_{rr}+(r+\frac{1}{r})R_r - a^2(1+\frac{1}{r^2}) R=0\\
\label{1.15}&&T_{\theta\theta}+ a^2T=0.
\end{eqnarray}
If we are interested in periodic solutions, we set $T(\theta)=A\cos(a\theta)+B\sin(a\theta).$ Thus
(\ref{1.14}) can be reduced to the normal form by the following functional change:
\begin{eqnarray*}
R \mapsto \widehat{R};~~~~ R(r)=\frac{1}{\sqrt{r}}\exp(-\frac{r^2}{4})\widehat{R}(r).
\end{eqnarray*}
Equation (\ref{1.14}) then becomes:
\begin{eqnarray} \label{1.16}
 \widehat{R}_{rr}+\left( -(1+a^2)+(\frac{1-4a^2}{4})\frac{1}{r^2} -\frac{r^2}{4}\right)\widehat{R}.
\end{eqnarray}
This equation can be reduced to the Whittaker equation \cite{Buc}.
Writting \\$\widehat{R}(r)= r^{-1/2}M(r^2/2)$, then the function $M(\varepsilon)$ sastifies the
 Whittaker equation:
\begin{eqnarray*}
 M(\varepsilon)=M_{\alpha, \mu/2}(\varepsilon);~
 \frac{d^2}{d\varepsilon^2}M_{\alpha, \mu/2}+(\frac{-1}{4}+\frac{\alpha}{\varepsilon}+
 \frac{1-\mu^2}{4\varepsilon^2})M_{\alpha, \mu/2}=0
\end{eqnarray*}
with $\alpha = \pm \frac{1}{2}(a^2+1),~~\mu=\pm a$.
This solution can be expressed in terms of  the confluent hypergeometric  function ${}_1F_1$
which leads to the solution of (\ref{1.14}) written as
\begin{eqnarray*}
R(r)=\frac{1}{2^{(a+1)/2}}e^{-r^2}r^a {}_1F_1(1+\frac{a}{2}+\frac{a^2}{2}, 1+a, r^2/2).
\end{eqnarray*}
If $a>0$, this solution is regular at the origin. We know that ${}_1F_1(m,m,x)=\exp(x)$.
Then, for $a=1$, we get $R(r)=r/2$, which is a solution of (\ref{1.14}). Nevertheless  this trivial
solution (and  related to the case $a=0$) leads to some problems because, if we set $T(\theta)=\cos\theta,~~ f(r,\theta)=r\cos\theta$,
 then $f_v =0$, $J= 0$.

\subsection{$E=\mathbb{R}^{1,1}$}
 The computations are similar to the preceding case. The equation (\ref{variation of the functional}) leads to
 \begin{eqnarray*}\label{variation_of_the_functional_when_E_R11}
 (1+\phi_x^2)\phi_{xx}-2\phi_x\phi_y\phi_{xy}-(1-\phi_y^2)\phi_{yy}=0.
\end{eqnarray*}
 The solutions of the form $\phi(x,y)=F(x)+F(y)$ can be written with a parametric representation:
\begin{eqnarray*}
 & &x=\frac{1}{\lambda}(\frac{p^3}{3}+p-c),~~~~y= \frac{1}{\lambda}(\frac{-q^3}{3}+q-c'),~~~\\
 & &\phi(x,y)=\frac{1}{4\lambda}(p^4+2p^2-q^4+2q^2)+cst.
\end{eqnarray*}

\section{Exponentially harmonic maps on Friedmann-Lema\^ itre universe}
Let M be a Friedmann-Lema\^ itre (FL) universe endowed with the following metric:
\begin{eqnarray}
ds^2=dt^2-R^2(t)\left( \frac{dr^2}{1-kr^2}+r^2(d\theta^2+\sin^2\theta d\phi^2)\right).
\end{eqnarray}
We  consider exponentially harmonic maps: $\phi:M \rightarrow \mathbb{R}$ which are,
for physical reasons explained above,  extremals of the functional $E'_\lambda$. They
satisfy a modified  version of equation (\ref{variation of the functional})  which is written in this case:
\begin{eqnarray}
 \ddot{\phi}(1+\lambda \dot{\phi}^2)+3\frac{\dot{R}}{R}\dot{\phi}=0,~~~~(\dot{\phi}=\phi_t,~~
 \dot{R}=R_t)
\end{eqnarray}
if we restrict ourselves to $\phi=\phi(t)$. This gives:
 \begin{eqnarray}
  R^3(t)=\frac{a}{\vert\dot{\phi}\vert}exp(-\frac{\lambda}{2}\dot{\phi}^2),~~~~~~~
  ( a: \mbox{positive constant}).
\end{eqnarray} 
Let us take, for example, $R(t)=R_o(\frac{t}{t_o})^{2/3}$ i.e.,  $M$ is  Euclidean FL-universe
(without cosmological constant) and $t_o=2/3h_o$ where $H_o$ is the present value of  the
 Hubble constant ($\dot{R}/R\vert_{t=t_o}=H_o$). We get the field equation
$$\vert \dot{\phi}\vert \exp(\frac{\lambda}{2}\dot{\phi}^2)=1/bt^2,~~~~
(b= R_o^3/at_o^2).$$
 i) When $t\rightarrow \infty$: this gives $\phi(t) \rightarrow$ constant.\\
 ii) When $t\rightarrow 0$ and $\lambda$ small: $\phi(t) \approx \phi_o \pm 1/bt$
which is not regular at $t=0$. See Figure1

When $\lambda \approx 0$ we recover the well-know theory of harmonic maps coupled with gravity
 studied by T. Hirschmann, R. Schimming \cite{HiS}, T. Hughes, T. Kato, J. Marden \cite{HKM} and
 Lemaire and A. J. Vanderwinden \cite{Van}.

\section{Exponentially harmonic maps on F-L universe coupled with gravity}
Let us start now with the following action, describing a gravitational field coupled with an
 exponentially scalar field:
\begin{eqnarray}\label{4.1}
S(\phi)= -\frac{1}{2k}\int\sqrt{-g}d^4x\left\lbrace \left( R-\exp(\frac{\lambda}{2}\partial_\alpha \phi\partial^\alpha\phi)
-\Lambda\right) +\mathcal{L}_{mat}\right\rbrace
\end{eqnarray}
 $K$ is a constant, $\Lambda$ is the cosmological constant, $\mathcal{L}_{mat}$ is the Lagrangian
 density for  matter and $\phi: M \rightarrow \mathbb{R}$ is a scalar field defined on a
 $4$-dimensional spacetime. The variation  of $S(\phi)$ leads to Einstein's equations:
 \begin{eqnarray*}
R_{\mu\nu}-\frac{1}{2}Rg_{\mu\nu}=\frac{1}{2}
\left\lbrace g_{\mu\nu}\left(- e^{\frac{\lambda}{2}\partial_\alpha \phi\partial^\alpha\phi}-\Lambda\right)
 -\lambda\partial_\mu\phi\partial_\nu\phi e^{\frac{\lambda}{2}\partial_\alpha \phi\partial^\alpha\phi}
 \right\rbrace + T_{\mu\nu}^{(mat)}
\end{eqnarray*}
where $T_{\mu\nu}^{(mat)}$ is the usual energy-momentum tensor for  matter. The variation with
respect to $\phi$ gives a field equation which is very similar to (\ref{variation of the functional}).
Let us assume that $\phi=\phi(t)$ and let $M$   be a Friedmann-Lema\^ itre universe. Then the field
equations can be written:
\begin{eqnarray}
 \label{4.2}
 &&3\left( \frac{\dot{R}}{R}\right)^2+3\frac{k}{R^2}=K\left( \rho-\frac{1}{2}e^{\frac{\lambda}{2}\dot{\phi}^2}( 1-\lambda\dot{\phi}^2)-\frac{\Lambda}{2}\right)\\
 \label{4.3}
 &&\left( \frac{\dot{R}}{R}\right) ^2+2\frac{\ddot{R}}{R}+\frac{k}{R^2}=K\left( -p-\frac{1}{2}e^{\frac{\lambda}{2}\dot{\phi}^2}-\frac{\Lambda}{2}\right)\\
 \label{4.4}
 &&\ddot{\phi}(1+\lambda \dot{\phi}^2)+3\frac{\dot{R}}{R}\dot{\phi}=0.
\end{eqnarray}
 Let us consider some particular cases.
\subsection{A flat F-L universe without matter}%{$\rho=p=0;~~~~~k=0$}
 \begin{eqnarray*}
&&3\left(\frac{\dot{R}}{R}\right)^2=K\left( \frac{1}{2}e^{\frac{\lambda}{2}\dot{\phi}^2}(- 1+\lambda\dot{\phi}^2)-\frac{\Lambda}{2}\right)\\
&&\left(\frac{\dot{R}}{R}\right)^2+2\frac{\ddot{R}}{R}=K\left( -\frac{1}{2}e^{\frac{\lambda}{2}\dot{\phi}^2}-\frac{\Lambda}{2}\right)\\
&&\ddot{\phi}(1+\lambda \dot{\phi}^2)+3\frac{\dot{R}}{R}\dot{\phi}=0.
\end{eqnarray*}
Let us define $y=\dot{\phi}$ and let $H=\frac{\dot{R}}{R}$ be the Hubble constant. We check that\\ $\dot{H}+H^2=\frac{\ddot{R}}{R}$, then the   equations above define a dynamical system:
\begin{eqnarray}
\label{4.5}
& &\dot{H}=-\frac{K\lambda}{4}y^2e^{\frac{\lambda}{2}y^2}\\
\label{4.6}
& &\dot{y}= -\frac{3 H y}{1+\lambda y^2}
\end{eqnarray}
 submitted to the constraint:
 \begin{eqnarray}
 \label{4.7}
 H^2= \frac{K}{6}\left( (-1+\lambda y^2)e^{\frac{\lambda}{2}y^2}-\Lambda\right).
\end{eqnarray}
Let us derive (\ref{4.7}) and, using (\ref{4.6}), let us compare the result with (\ref{4.5}). We see that the preceding equations are compatible.
We are thus led to solve the following equation:

\begin{eqnarray}\label{4.8}
 \frac{\ddot{y}}{y}(1+\lambda y^2)- \left( \frac{\dot{y}}{y}\right)^2(1-\lambda y^2)
   =\frac{3}{4}K\lambda y^2e^{\frac{\lambda}{2}y^2} .
\end{eqnarray}
For very small values of $\lambda$, this becomes:
 \begin{eqnarray}\label{4.9}
  \frac{\ddot{y}}{y}-\left(\frac{\dot{y}}{y}\right)^2\approx 0.
\end{eqnarray}
 This gives solutions:
 \begin{eqnarray*}
 \phi(t) = ce^{at}+ e; ~~~~~~R(t)=be^{Ht}
\end{eqnarray*}
where $a,c$ are constants which the same sign, $b$ a positive constant and $e$ is an arbitrary constant.
$H=-\frac{a}{3}$  ($H$ is a first integral when $\lambda=0$).
The equation (\ref{4.7}) allows to write $\Lambda=-1-\frac{2a^2}{3K}$.   See Figure 2

It is interesting to note here that the coupling of $\phi$ with the gravitational field can make $\phi$
regular at $t=0$ in this Euclidean case, which was not the case in the uncoupled situation
(see section3).

 Let us return to the case where $\lambda$ is an arbitrary constant. We define $z=\dot{y}/y$, equation
 (\ref{4.8}) leads to
\begin{eqnarray*}
\dot{y}&=& zy\\
\dot{z}&=& \frac{2\lambda y^2}{1+\lambda y^2}(\frac{3K}{8}e^{\frac{\lambda }{2}y^2}-z^2).
\end{eqnarray*}
which is a dynamical system with a fixed point $(y, z) = (0, 0)$. If we put $u=y^2$ we get:
\begin{eqnarray}
\label{4.10} \dot{u} &=& 2 zu\\
\label{4.11} \dot{z}&=& \frac{2\lambda u}{1+\lambda u}(\frac{3K}{8}e^{\frac{\lambda }{2}u}-z^2).
\end{eqnarray} 
The Hubble constant can be deduced from the numerical integration of this dynamical system.
 From (\ref{4.7}) we get:
\begin{eqnarray}
H = \pm \left[- \frac{K}{6} \left( (1-\lambda u) e^{\frac{\lambda}{2}u}+\Lambda \right) \right] ^{1/2}.
\end{eqnarray}
Around $u\approx 0$ (i.e. $\phi=$ constant) we can neglect  terms of order greater or equal to $2$.
Using (\ref{4.10}) - (\ref{4.11}) we have at  first order:
\begin{eqnarray}
u\approx -\frac{1}{2\lambda} ln\left( \vert \lambda\vert(-\frac{3K}{4}+2 z^2)\right) +u_o.
\end{eqnarray} 
 As $u=y^2$, we have to suppose that (for $\lambda > 0$): $ -\frac{3K}{4}+2 z^2 < \frac{1}{\lambda}$
 when $u_o=0$.

\subsection{A curved F-L universe without matter}%{$\rho=p=0,~~~~ k \neq 0$}
Let us derive (\ref{4.4}). Using (\ref{4.2}) and (\ref{4.3})  with $y=\phi$,we find  the following equation:
\begin{eqnarray}\label{4.12}
 12(1+\lambda y^2)\frac{\ddot{y}}{y}+4(\lambda y^2-\lambda^2 y^4-4)\left( \frac{\dot{y}}{y}\right)^2
\nonumber\\-3K(\lambda y^2+2)e^{\frac{\lambda}{2}y^2}-6K\Lambda=0.
\end{eqnarray}
Equation (\ref{4.12}) can be solved numerically using the  variable change: $ u= y^2$.
Then (\ref{4.12}) becomes:
\begin{eqnarray}\label{4.13}
 6(1+\lambda u)\frac{\ddot{u}}{u}-(\lambda^2 u^2-2\lambda u+7 )\left( \frac{\dot{u}}{u}\right)^2
-3K(\lambda u +2)e^{\frac{\lambda}{2}u}-6K\Lambda=0.
\end{eqnarray}
Making the substitution $z=\frac{1}{2}\frac{\dot{u}}{u}=\frac{\dot{y}}{y}$, (\ref{4.13}) can be written as.
\begin{eqnarray}
 \label{4.15} \dot{u}&=&2 z u\\
 \label{4.16} \dot{z}&=&\left[ \frac{1}{3}(\lambda^2 u^2-4\lambda u +1)z^2+\frac{1}{4}K(2+\lambda u)e^{\frac{\lambda}{2}u}+ \frac{1}{2}K\Lambda\right] \frac{1}{(\lambda+1)}
\end{eqnarray}
wich can be studied numerically.

Using (\ref{4.4}), we can determine $R(t)$
\begin{eqnarray} \label{4.14}
\frac{1}{R^3}= \alpha \dot{\phi} \exp{\frac{\lambda}{2}\dot{\phi}^2}.
\end{eqnarray} 
If $\lambda \approx 0$, equation  (\ref{4.12}) leads to (\ref{4.9})
with $\frac{\ddot{y}}{y}\approx \frac{\dot{y}}{y}=-\frac{3}{2}K(1+\Lambda)$ if we assume that $k=0$. If $ k\neq 0$, the case $\lambda \approx 0$ can be studied from:
\begin{eqnarray*}
 \frac{\ddot{y}}{y}\approx \frac{4}{3}\frac{\dot{y}}{y}+\frac{K}{2}(\Lambda+1).
\end{eqnarray*}
With   the above notation this gives
\begin{eqnarray}
\dot{z} \approx \frac{1}{3}z^2 +\frac{K}{2}(\Lambda+1).
\end{eqnarray} 
 The case  $\Lambda=-1$ is obvious. When $\Lambda > -1$, we get:
\begin{eqnarray*}
z(t) &\approx& \left( \frac{3K}{2}(\Lambda+1)\right)^{1/2}\tan\left[ \frac{1}{3} \left(\frac{3K}{2}(\Lambda+1)\right)^{1/2}(t-t_o)\right]\\
%\mbox{or}~~~~~& &\\
\end{eqnarray*}
or
\begin{eqnarray*}
y(t)&\approx& \left\lbrace \cos\left[\frac{1}{3} \left(\frac{3K}{2}(\Lambda+1)\right)^{1/2}(t-t_o)\right]\right\rbrace^{-3}+c
\end{eqnarray*}
where $c$ is a constant. We have thus:
\begin{eqnarray}
\phi(t)&=& \exp{\left\lbrace 3\left( \frac{3K}{2}(\Lambda+1)\right)^{-1/2}\frac{\sin\left[\frac{1}{3} \left(\frac{3K}{2}(\Lambda+1)\right)^{1/2}(t-t_o)\right] }{2\cos^2\left[\frac{1}{3} \left(\frac{3K}{2}(\Lambda+1)\right)^{1/2}(t-t_o)\right] }\right\rbrace }\nonumber\\
&\times& \left[ \tan \left( \frac{\pi}{4}+\frac{1}{6} \left(\frac{3K}{2}(t-t_o)\right)^{1/2}\right) \right] ^{\frac{3}{2}\left(\frac{3K}{2}(\Lambda+1)\right)^{-1/2}}\exp{(ct)}.
\end{eqnarray}
When $\Lambda <-1$ we have:
\begin{eqnarray}
 \phi(t)&=& \exp\left\lbrace -3\left( \frac{3K}{2}(\Lambda+1)\right)^{-1/2}\frac{\sinh\left[\frac{-1}{3} \left(\frac{3K}{2}(\Lambda+1)\right)^{1/2}(t-t_o)\right] }{2\cosh^2\left[\frac{-1}{3} \left(\frac{3K}{2}(\Lambda+1)\right)^{1/2}(t-t_o)\right] }\right.\nonumber\\
&+&\left. \frac{1}{2} \tan^{-1} \left[ \sinh\frac{-1}{3}  \left(\frac{3K}{2}(-1-\Lambda)\right)^{1/2}(t-t_o) \right] \right\rbrace  \exp{(ct)}.
\end{eqnarray}

%\begin{eqnarray}
%\phi(t)&=& \exp \left\lbrace  -3\left( \frac{3K}{2}(\Lambda+1)\right)^{-1/2}
%\frac{\sinh\left[\frac{-1}{3} \left(\frac{3K}{2}(\Lambda+1)\right)^{1/2}(t-t_o)\right] }{2\cosh^2\left[\frac{-1}{3} \left(\frac{3K}{2}(\Lambda+1)\right)^{1/2}(t-t_o)\right] } \right.\nonumber\\
%\left. &+& \frac{1}{2} \tan^{-1} \left[ \sinh\frac{-1}{3}  \left(\frac{3K}{2}(-1-\Lambda)\right)^{1/2}(t-t_o) \right] \right\rbrace}    \exp{(ct)}.
%\end{eqnarray}

\subsection{A F-L universe with matter: $p=\omega\rho$}%{$\rho \neq 0$,  $p=\omega\rho$}
Let us multiply (\ref{4.2}) by $\omega$ and let us add (\ref{4.2}) and (\ref{4.3}). We get:
\begin{eqnarray}
& &\left( \frac{\dot{R}}{R}\right)^2(3\omega+1)+\frac{2\ddot{R}}{R}+\frac{k}{R^2}(3\omega+1)\nonumber\\
& &+\frac{K}{2}e^{\frac{\lambda\dot{\phi}^2}{2}}(1+\omega-\lambda\omega\dot{\phi}^2)+\frac{K}{2}\Lambda(\omega+1)=0.
\end{eqnarray} 
Using (\ref{4.4}), we find:
\begin{eqnarray}
 \frac{\ddot{R}}{R} -\frac{\dot{\ddot{\phi}}}{3\dot{\phi}}(1+\lambda\phi^2)+\frac{\ddot{\phi}^2}{9\dot{\phi}^2}(\lambda^2\dot{\phi}^4-\lambda\dot{\phi}^2+4).
\end{eqnarray} 

Then, by (\ref{4.2}) and using the preceding notation, we can write the the equation:
\begin{eqnarray}\label{4.17}
\ddot{y}&=&\frac{\dot{y}^2}{y}\left( \frac{(3+\omega)+2\lambda\omega y^2+\lambda^2(1+\omega)y^4}{2(1+\lambda y^2)}\right) +\frac{3}{4}\frac{y K\Lambda(\omega +1)}{1+\lambda y^2}\nonumber\\
&+&\frac{3}{4}Ke^{\frac{\lambda}{2}y^2}\frac{(1+w)y-\lambda\omega y^3}{1+\lambda y^2}+\frac{3}{2}}e^{\frac{\lambda}{3}y^2}{\alpha^{2/3} y^{5/3}\lambda^{1/3}\frac{k(3\omega+1)}{1+\lambda y^2}
\end{eqnarray}
which is nothing but (\ref{4.12}) when $\omega=0$ and $\rho=0$; $\rho=0$ implies $\alpha =0$
(see (\ref{4.14})). Using the conservation law $\nabla_\mu T^{\mu\nu}_{mat} = 0$ and equation(\ref{4.4}),
we find:

\begin{eqnarray}
\label{divt0}
 \frac{\dot{\rho}}{\rho(1+\omega)}=-3\frac{\dot{R}}{R}=\frac{(1+\lambda)\dot{y}}{y}
\end{eqnarray}
  %Numerical solutions of Eq. (\ref{4.17})  of has been represented in Figs. 3, 4 and 5
An example of numerical solutions  is given in Fig.3, 4 and 5.

In Fig.3, we can see the typical behaviour of a Friedmann-Lema\^ itre universe whose expansion is driven by exponentially
harmonic maps and matter or radiation. This evolution know three different ages:
first, the universe behaves in  the usual way under the presence of matter or radiation, then its characteristic deceleration
stops and finally, when the exponentially harmonic map  $\phi$ becomes too small, the expansion accelerates (see Eq.(\ref{4.14})).
But this is also the typical bahaviour of a universe that is dominated eventually by a positive cosmological constant or something
that acts like it, such as a potential related to a scalar field (quintessence). Nevertheless, the careful reader has already noticed
that this particular behaviour is only due to the exponentially harmonic feature of the scalar field $\phi$ and its particular initial
conditions $\dot{\phi_0}$ and $\ddot{\phi_0}$. This proves again that these interesting features, suggested among others by the
observations of type Ia supernovae, arise naturally in the context of exponentially harmonic maps without requiring an adhoc
mechanism of inflation.

In Fig.4, we have represented the evolution of the energy densities of matter, radiation and the scalar field coupled to it. In
the very early ages of the universe\footnote{It is interesting to note that here there is a huge difference in the time scales of
Figs. 3 and 4, about 7 orders of magnitude.}, the dominant contribution to the total energy is due to $\phi$. But the universe
expands just as if there was only usual matter and radiation.   As this expansion goes on, the energy density of the scalar
field $\rho_\phi$ goes below the density of matter and radiation, making the expansion accelerating when the field $\phi$
become tiny (see (\ref{4.14})). This would correspond to the well known   \textit{ ``cosmic coincidence''}. Note that we have chosen the
arbitrary constants arising in the integration of Eq.(\ref{divt0}) in order to get this peculiar model of expansion. The present
study is purely qualitative,  more  work is under way to construct a quantitative model that could be compared, for example,
with those of quintessence.

In Fig.5, we have represented the evolution of the scalar field $\phi$ in order to satisfy Eq.(\ref{4.17}) both in the presence of
matter ($\omega=0$) and radiation ($\omega=1/3$).

% It is interesting to note here that the solution corresponding to fig5.  shows an accelerating phase which
%is qualitatively similar to the universe's model confirmed recently by observations on distant supernovae
%of Ia type. This behaviour is very different from the case in fig4 with $\rho=0$ and $k\neq 0$ (see section 4.2)).
%It worth nothing that the value of the field $\phi$ is nearly constant far very recent time and its
%characteristic variation occurs only at the beginning of the universe.

\section{Conlusions}
The main interest of the use of exponentially harmonic maps is the fact  that the Lagrangian (\ref{4.1})
is a generalisation of bosonic string Lagrangian (with only a dilatonic field) written in Einstein
frame \cite{LWC}. Indeed, if $\lambda\approx 0$, the Lagrangian (\ref{4.1}) tends to this bosonic string
Lagrangian, if $\Lambda=-1-\Lambda_o$, where $\Lambda_o$ is the small cosmological constant.
But mathematically, this limit $\lambda \rightarrow 0$ is highly problematic due to the fact
that unlike usual harmonic maps, exponentially harmonic maps have no good invariance property under
homothetic changes of the field,  as it was shown in the introduction.

An interesting  open question  is whether solutions of equations deriving from (\ref{4.1})
admit duality symmetries. We are in the process of studying this question.

\newpage
\begin{figure}
\begin{center}
\includegraphics[scale=0.4,angle=0]{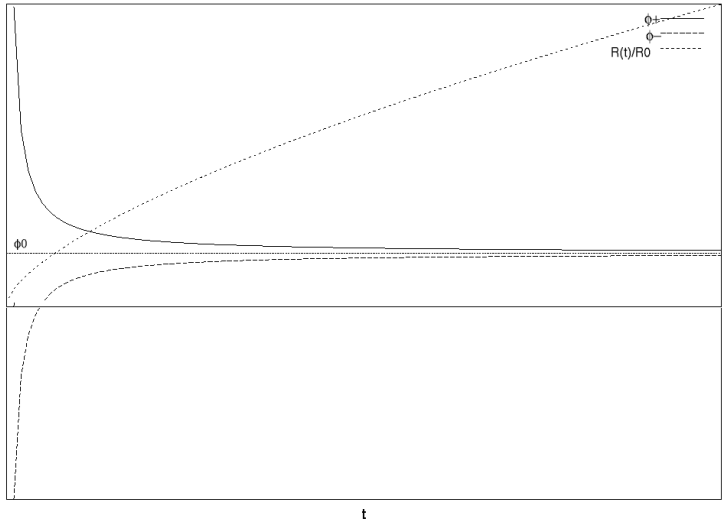}
\caption{\footnotesize \textsl{Behaviour of the expansion factor $R(t)$ and the exponentially harmonic map $\phi(t)$ on 
an euclidean Friedmann-Lema\^ itre spacetime ($\lambda\approx 0$ ; $\Lambda=0$ and $\phi_0 >0$).
} \normalsize}
\label{fig1}
\end{center}
\end{figure}

\begin{figure}
\begin{center}
\includegraphics[scale=0.25,angle=0]{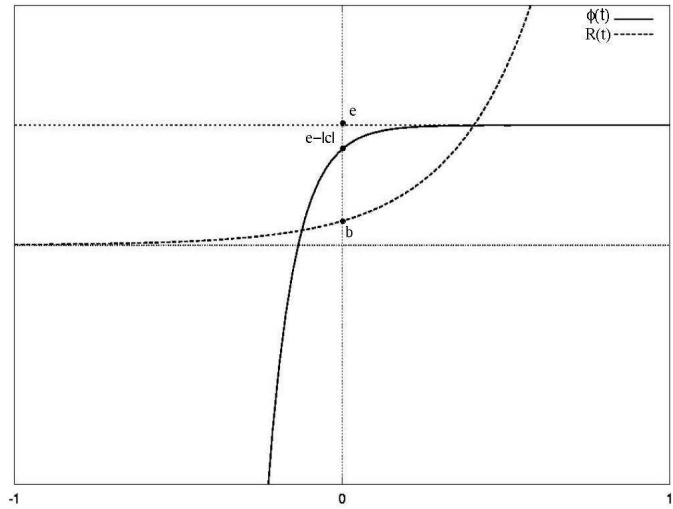}
\caption{\footnotesize \textsl{Behaviour of the expansion factor $R(t)$ and the exponentially harmonic map $\phi(t)$ on 
an euclidean Friedmann-Lema\^ itre universe in the Einstein frame. ($\lambda\approx 0$ ; $\Lambda=0$; $e>|c|+b$ and $a=-\left(\frac{3\kappa}{2}\right)^{1/2}$).
} \normalsize}
\label{fig2}
\end{center}
\end{figure}

\begin{figure}
\begin{center}
\includegraphics[scale=0.4,angle=0]{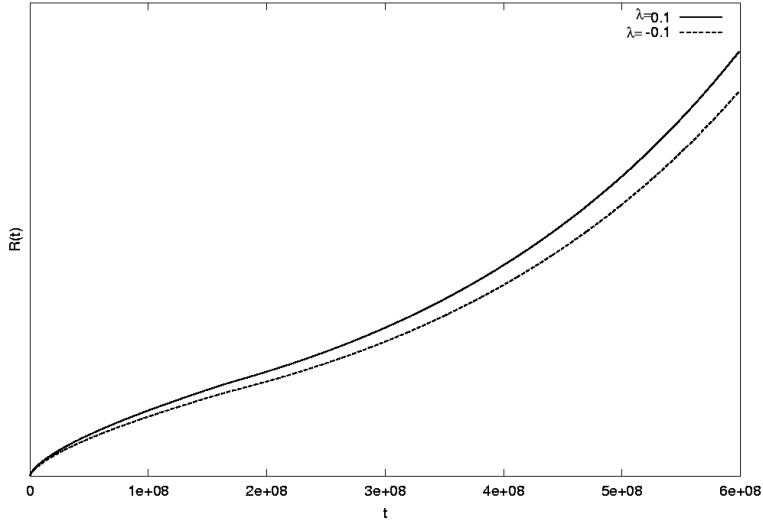}
\caption{\footnotesize \textsl{Expansion factor computed from Eqs. $49$ and $43$ illustrating the behaviour of an hypothetic universe
filled with matter ($\omega=0$) and an enxponentially harmonic map. ($\dot{\phi_0}=1$, $\ddot{\phi_0}=-5$, $\omega=0$).
} \normalsize}
\label{fig3}
\end{center}
\end{figure}

\begin{figure}
\begin{center}
\includegraphics[scale=0.25,angle=0]{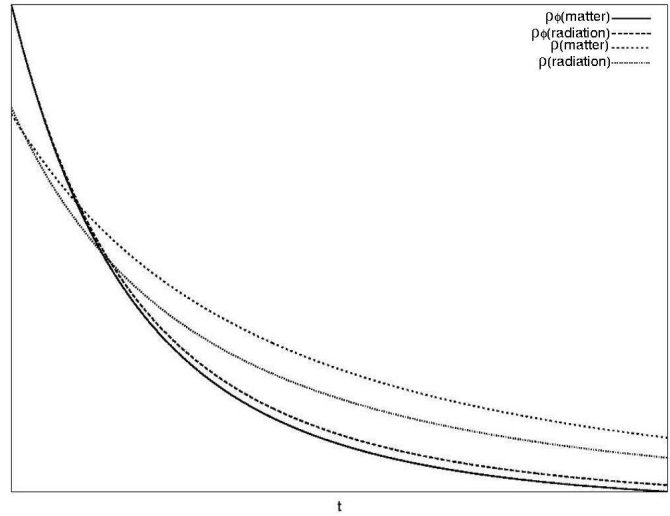}
\caption{\footnotesize \textsl{Evolution of energy densities of matter, radiation and the coupled scalar field. The arbitrary constants arising in the integration of Eq.\ref{divt0} have been chosen in order to draw all the curves in the same plot.
($\dot{\phi_0}=1$, $\ddot{\phi_0}=-5$, $\lambda=0.1$).
} \normalsize}
\label{fig4}
\end{center}
\end{figure}

\begin{figure}
\begin{center}
\includegraphics[scale=0.3,angle=0]{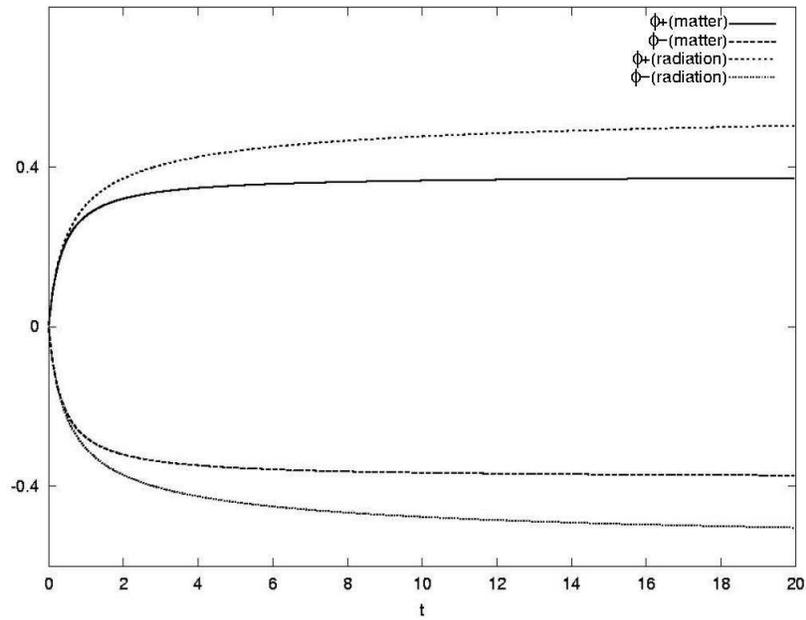}
\caption{\footnotesize \textsl{Exponentially harmonic maps $\phi=\pm\sqrt u$ solution of Eq.49, for $\omega=0$ (dust-dominated universe) and $\omega=1/3$ (radiation-dominated universe).($\dot{\phi_0}=1$, $\ddot{\phi_0}=-5$, $\lambda=0.1$).} \normalsize}
\label{fig5}
\end{center}
\end{figure}

\end{document}